\documentclass[prd,12pt]{article}

\usepackage{tikz}

\definecolor{DarkGreen}{rgb}{0,0.6,0}
\usepackage{amsmath,amssymb,graphicx,multirow,xspace,slashed}
\usepackage[colorlinks=true,linktocpage=true,linkcolor=DarkGreen,citecolor=red,urlcolor=blue]{hyperref}
\usepackage[compress,numbers]{natbib}
\usepackage{subfigure, placeins}
\usepackage{booktabs}
\usepackage{blindtext, rotating}
\usepackage{afterpage}
\usepackage{enumitem}
\usepackage{ marvosym }
\usepackage{authblk} 
\usepackage{verbatim}
\usepackage{soul} 
\usepackage[normalem]{ulem}
\usepackage{pifont}
\usepackage[usestackEOL]{stackengine}
\usepackage{subfiles}
\newcommand\sq{\framebox(10,10){}\kern\fboxrule}

\setstackgap{S}{\fboxrule}

\usepackage[utf8]{inputenc}

\usepackage{floatrow}
\newfloatcommand{capbtabbox}{table}[][\FBwidth]

\usepackage[font=footnotesize,labelfont=bf]{caption}

\usepackage{lineno}

\usepackage{ulem}

\allowdisplaybreaks

\long\def\symbolfootnote[#1]#2{\begingroup%
\def\thefootnote{\fnsymbol{footnote}}\footnote[#1]{#2}\endgroup}

\newcommand{\newc}{\newcommand}
\newc{\gsim}{\lower.7ex\hbox{$\;\stackrel{\textstyle>}{\sim}\;$}}
\newc{\lsim}{\lower.7ex\hbox{$\;\stackrel{\textstyle<}{\sim}\;$}}
\newc{\gev}{\,{\rm GeV}}
\newc{\mev}{\,{\rm MeV}}
\newc{\ev}{\,{\rm eV}}
\newc{\kev}{\,{\rm keV}}
\newc{\tev}{\,{\rm TeV}}
\newc{\MHT}{$H_T^{\text{miss}}$}
\newc{\MET}{$\slashed{E}_T$}
\newc{\MTT}{$M_{T2}$}

\def\ln{\mathop{\rm ln}}

\def\Tr{\mathop{\rm Tr}}

\newc{\mz}{M_Z}
\newc{\mpl}{M_*}
\newc{\mw}{m_{\rm weak}}
\newc{\nr}[1]{N^c_R{}_{#1}}

\def\beq{\begin{equation}}
\def\eeq{\end{equation}}
\newcommand{\bea}{\begin{eqnarray}\begin{aligned}}
\newcommand{\eea}{\end{aligned}\end{eqnarray}}
\def\bitem{\begin{itemize}}
\def\eitem{\end{itemize}}
\newcommand{\nn}{\nonumber}

\definecolor{darkgreen}{rgb}{0,0.5,0}
\definecolor{goodyellow}{rgb}{0.9,0.7,0}

\numberwithin{equation}{section}

\newcommand\fverb{\setbox\fverbbox=\hbox\bgroup\verb}

\newbox\fverbbox

\begin{document}
\baselineskip 0.6cm

\begin{titlepage}

\thispagestyle{empty}

\begin{center}

\vskip 0.1cm

\hspace{4.8in}

\vspace{1.3cm}

{\Large \bf Quantum Entanglement and the Thermal Hadron}

\vskip 0.5cm

\vskip 0.5cm

\vskip 1.0cm
{\large Pouya Asadi$^{1}$, Varun Vaidya$^{2}$}
\vskip 1.0cm
\textit{${}^1$Institute for Fundamental Science and Department of Physics, \\ University of Oregon, Eugene, OR 97403, USA} \\\vskip0.35em
{\it $^2$ Department of Physics, University of South Dakota, \\ Vermillion, SD 57069, USA.\\}
\vskip 0.3cm

\end{center}

\vskip 0.6cm

\begin{abstract}
This paper tests how effectively the  bound states of strongly interacting gauge theories are amenable to an emergent description as a thermal ensemble. This description can be derived from a conjectured minimum free energy principle, with the entanglement entropy of two-parton subsystems playing the role of thermodynamic entropy. This allows us to calculate  the ground state hadron spectrum and wavefunction over a wide range of parton masses without solving the Schr\"{o}dinger equation. 
We carry out this analysis for certain illustrative models in 1+1 dimensions and discuss prospects for higher dimensions.

\end{abstract}

\flushbottom

\end{titlepage}

\setcounter{page}{1}

\tableofcontents

\vskip 1cm

\section{Introduction}
\label{sec:intro}

The holy grail of nuclear physics is  to predict the hadron spectrum and wavefunction from first principles. While Quantum Chromodynamics (QCD) accurately explains the underlying UV theory, it can not answer these questions, primarily due to lack of any systematic expansion parameter in the IR. The most natural candidate for such an expansion is around a free theory. However, the running of this coupling to large values in the infrared, thanks to asymptotic freedom, disqualifies a perturbative expansion around the non-interacting theory. 

As a result, many alternative novel proposals for studying strongly interacting theories have been put forward, e.g. see Refs.~\cite{Skyrme:1961vq,Skyrme:1962vh,Schwinger:1962tp,tHooft:1973alw,Chodos:1974je,Chodos:1974pn,tHooft:1974pnl,DeGrand:1975cf,Callan:1977gz,Shifman:1978bx,Witten:1979kh,Witten:1983tx,Adkins:1983ya,Komargodski:2018odf}. One possibility is to map the strongly interacting theory to a weakly coupled one via dualities between two local theories, e.g. between the Sine-Gordon and the Thirring model in 1+1 dimension (1+1D) \cite{Coleman:1974bu} or the AdS/CFT correspondence \cite{Maldacena:1997re} (see also \cite{Witten:1998qj}). 
However, such dualities are exceptions rather than the rule \cite{Cotler:2017abq}, and so far no dual theory to QCD has been discovered. 

A renowned proposal \cite{tHooft:1973alw} is to use the inverse of number of colors as a small expansion parameter, i.e. the large $N$ expansion.  This was shown to simplify the calculation considerably in 1+1D models \cite{tHooft:1974pnl,Witten:1979kh}, but ultimately was unsuccessful in solving the problem in higher dimensional gauge theories and has been only used for simplifying certain calculations in higher dimensions such as the BK equation \cite{Balitsky:1995ub,Kovchegov:2012mbw}.

Nevertheless, exactly solvable models in lower dimensions have enormous value as a testbed for assaying new alternative approaches to non-perturbative physics. 
Models in 1+1D have the advantage of fewer degrees of freedom, e.g. no spin or transverse modes, and that the renormalization of the structure functions is power suppressed. 
These simplifications, along with certain exact dualities, have allowed analytic solutions to be obtained for some models in 1+1D, e.g. see Refs.~\cite{Schwinger:1962tp,tHooft:1974pnl}. 
The hope is that the nature of the solution in such exactly solvable models can give us a clue about some simple emergent principle that results from strong interactions between the free theory degrees of freedom. 
Such a sought-after principle ought to be general enough that it is not tied to the special properties of the model, and therefore be portable to theories in higher dimensions. 
Nevertheless, for any proposed principle to ultimately have value, we need to first be able to demonstrate its validity for $distinct$ models in 1+1D.

We currently gather information about hadron structures mainly through scattering experiments, such as Deep Inelsatic Scattering (DIS), in which the internal structure of a hadron is accessed through universal, yet non-perturbative, functions such as parton distribution functions (PDFs), transverse momentum PDFs (TMDPDFs), generalized parton distributions (GPDs), etc., each carrying less information than the full wavefunction. 
These functions do not have any analytical solution and determining them is almost completely reliant on experimental results and measurements. (Even first principle numerical calculations of PDFs is a very recent development \cite{Ji:2013dva,Xiong:2013bka,Ji:2014gla,Ma:2014jla,Chen:2017mzz,Ma:2017pxb,Zhang:2018nsy}.) 
So a natural goal of any non-perturbative approach would be to reproduce the form of the PDFs by comparison with data or numerical computations.

In this paper, we take the view that it is worthwhile searching for a physically motivated principle that might arise out of the complex non-perturbative dynamics of strongly coupled theories. 
If we think of hadrons as containing many partons that are interacting within a limited phase space, it is reasonable to assume that, thanks to numerous and strong interactions, this many-parton system traverses every corner of its phase space, i.e. follows the ergodic hypothesis, and eventually approaches an equilibrium distribution. 
While the ergodic hypothesis may not hold for all Hamiltonian systems, we want to test the regime of its applicability for strongly coupled theories. 
As taught in introductory statistical mechanics, for systems reaching an equilibrium, macroscopic properties are derived from the maximum entropy principle, i.e. we can derive the well-known (micro/grand-) canonical ensembles from maximizing the Shannon entropy of the ensemble subject to various conservation constraints. 
This underscores the (quantum) information theory roots of the statistical mechanics, see Ref.~\cite{PhysRev.106.620} for a seminal work underlining this connection.  
More generally, the maximum entropy principle can be considered a special case of the minimum free energy principle in studying thermodynamics systems. 

Following the analogy between such conventional in-equilibrium systems and hadrons, statistical and quantum information principles have already been used in the literature to model various properties of hadrons with varying degrees of success \cite{Wang:2014lua,Han:2020vjp}. Furthermore, there has been enormous interest in trying to uncover the entanglement structure of hadrons from scattering experiments \cite{Simak:1988qp,Kutak:2011rb,Kharzeev:2017qzs,Shuryak:2017phz,Baker:2017wtt,Hagiwara:2017uaz,Liu:2018gae,Han:2018wsw,Tu:2019ouv,Castorina:2020cro,Iskander:2020rkb,Kharzeev:2021yyf,Baty:2021ugw,Dvali:2021ooc,Kharzeev:2021nzh,Zhang:2021hra,Hentschinski:2021aux,Dumitru:2022tud,Wang:2022noa,Benito-Calvino:2022kqa} (see also Refs.~\cite{Witten:1998zw,Aharony:2003sx,Klebanov:2007ws,Bah:2007kcs,Fujita:2008zv,Kol:2014nqa,Arefeva:2020uec,Jokela:2020wgs,daRocha:2021ntm} for use of entanglement entropy in studying phase transitions in confining theories).

In this paper, we treat the parton inside a hadron as an open quantum system interacting strongly with its environment. 
For such a system the von Neumann entropy of the reduced density matrix is just the entanglement entropy of the parton degree of freedom with its environment. 
We formulate \textit{a version of the minimum free energy principle for a hadron using the entanglement entropy of its two-parton subsystems, from which both its mass and its PDF can be approximately predicted over a wide range of parton masses.}  
In the limit of massless partons, minimizing this free energy function is equivalent to maximizing the entanglement entropy between every pairs of partons. 
Said differently, we find that ground state hadrons minimize a free energy function.

Given the form of the resulting ansatz, we are effectively testing how accurately can a hadron be represented as a thermal gas of partons. 
We will investigate the universality of this principle by applying it to different models in 1+1D, both for massless and massive partons.
Using this ansatz we can make fairly accurate predictions for the massless case, as well as for moderate to large parton masses.

We have less success for the  case of small but finite parton mass in these models, which is still an on-going area of research, see Ref.~\cite{Anand:2021qnd} for a recent study; we will point out how, with our proposed principle, further progress can be made in this limit as well.

The outline of the rest of the paper is as follows.  In the next section we will introduce our conjecture on use of the minimum free energy principle for studying hadrons. We also review basics of PDFs in 1+1D and derive an ansatz for it from our proposed first principle. In Sec.~\ref{sec:meson} (Sec.~\ref{sec:baryon}) we will use our conjecture to study mesonic (baryonic) bound states in a few toy models in 1+1D. We conclude in Sec.~\ref{sec:conclusion}. 
\section{Conjecturing an Alternative First Principle}
\label{sec:alternative}

Properties of various many-body systems are studied using maximum entropy principle or minimum free energy. 
To properly formulate and use maximum entropy principle, one needs to know the (Hilbert) space under study. 
Then one needs to decompose this (Hilbert) space into a system and an environment, trace out the environment, and calculate an entropy ansatz using the resulting reduced density matrix of the system; usually the Shannon (von Neumann) entropy of a subsystem of the system under study is used for classical (quantum) many-body systems, while many other entropy ansatzes exist as well.  
When in equilibrium, by definition, the state of the system is such that it maximizes this entropy subject to some constraints, e.g. conservation of average energy or average particle number or other quantum numbers of the system. 

The choice of the basis is decided by the nature of the interaction between the system and the environment. Here we are interested in the bound state of strongly coupled $local$ quantum field theories. Since entropy is intuitively a measure of mixing in a system, we use free theory momentum eigenstates, which are strongly mixed via \textit{local} interactions. This choice of basis is fortuitously suitable for our study of PDFs, which are defined as the momentum distribution of partons. We will study these properties in a few exactly-solvable models in 1+1D. 
In keeping with our goal to make connection with the PDF, we work in the Infinite Momentum Frame (IMF) where the bound state is boosted to a very large light-cone momentum ($P^-$) and small $P^+$ component such that $P^+P^- = M_{\text{hadron}}^2$. 
In the IMF, $P^+$ operator is identical to the Hamiltonian \cite{Brodsky:1997de}.

Theories in 1+1D enjoy a host of simplifications, foremost among them is the restriction of the bound state Fock space to valence partons.\footnote{The sea quark contributions are still non-zero, but are very suppressed, e.g. see Ref.~\cite{Hornbostel:1988fb}. }
In particular, the full Hilbert space of a meson (baryon) can be decomposed into the Fock space of a single quark and a single anti-quark ($N$ identical quarks). Each single parton space is spanned by states with different fractional longitudinal  momenta in the light-cone frame; the contribution from higher Fock states is suppressed (at least) for the lightest bound states \cite{Hornbostel:1988fb}. 
Furthermore, gluons are not propagating degrees of freedom and the PDF is independent of scale \cite{Kroger:1998se}. 
This follows from the fact that the gauge coupling $g$ is dimensionful and all radiative corrections are power suppressed by $g^2/Q^2$, where $Q$ is the high energy scale at which the hadron is probed in a DIS experiment. 
Finally, there is no notion of spin in 1+1D, thus we can only talk about unpolarized PDFs.

For asymptotically free theories in 3+1D, while there are no dimensionful couplings in the UV, an intrinsic scale is generated in the IR via dimensional transmutation. The internal dynamics of the bound state then depend heavily on how this scale, $\Lambda$, compares with the bare parton masses.
In 1+1D, the role of $\Lambda$ is played by the gauge coupling $g$, which is now dimensionful. 
In this case, two extreme limits arise. 

In one limit, when parton masses are zero, we expect the dynamics to be dominated by the potential term of the Hamiltonian, since the $P^+$ vanishes in the IMF in this limit. It is natural to surmise that strong interactions between partons maximally entangles them; thus, we conjecture that
\begin{itemize}
    \item In the limit of massless partons in a confining gauge theory, internal dynamics of the system maximizes the entanglement entropy between the partons of any two-parton subsystem carrying a fixed total momentum. We will subsequently give this statement a precise mathematical form.
\end{itemize}

In the other limit, when parton masses are much greater than $\Lambda$, the free part of the Hamiltonian dominates and we expect the bound state to be made up of quasi-free partons that are interacting relatively weakly. 
In order to interpolate between these two extremes, we therefore propose our main conjecture:
\begin{itemize}
    \item The ground state hadron of a confining gauge theory minimizes the free energy of its two-parton subsystem with fixed total momentum, defined as 
\bea 
F= E - T S.
\label{eq:defF}
\eea
Here $T$ is an intrinsic temperature to be determined variationally, $E$ is the expectation value of free parton Hamiltonian in the light-cone frame ($E \rightarrow 0$ as $m_q \rightarrow 0$ with $m_q$ denoting the parton's mass \cite{tHooft:1974pnl}), and $S$ is an entropy function capturing the entanglement of pairs of partons in a hadron, see Secs.~\ref{subsec:pdf-intro-mesons}--\ref{subsec:pdf-intro-baryons} for further details about the entanglement entropy used. 
\end{itemize}

We can think of our proposal as a standard variational method for finding the wavefunction (eigenfunction) and the mass (eigenvalue) of hadrons Hamiltonian, with $T$ being the variation parameter. Our innovation is to use the ansatz that minimizes the free energy of two-quarks subsystems, as defined in Eq.~\eqref{eq:defF}. 
As we will see, effectively we are proposing the interaction of a single parton with its environment is  simulated as a thermal bath, which is already proposed in Ref.~\cite{Kharzeev:2017qzs} in a different context. 
We find that this functional form accurately predicts the wavefunction and the mass of the ground state hadron in certain limits of the parameter space, for various toy models in 1+1D with one flavor of fundamental fermions.\footnote{As will be clear momentarily, all the details of the model appear in the potential term in the Hamiltonian, and not in the free energy expression. Thus, we expect the functional form of the wavefunctions to be the same if the fermion representations are changed, yet the hadron intrinsic temperature and mass will be modified.} 
These models include 1+1D QED, aka Schwinger model \cite{Schwinger:1962tp}, 1+1 SU($N$) with $N \rightarrow \infty$ \cite{tHooft:1974pnl} as well as 1+1 SU($N$) with finite $N$, i.e. 't Hooft model. 
Analytic or numerical results for mass and PDF of the ground state of these models exist in the literature, e.g. see Refs.~\cite{Schwinger:1962tp,tHooft:1974pnl,Hornbostel:1988fb,Kroger:1998se,Jia:2017uul}.

In the following two sections, we apply our conjecture to mesonic (Sec.~\ref{subsec:pdf-intro-mesons}) or baryonic (Sec.~\ref{subsec:pdf-intro-baryons}) bound states. The discussion of Sec.~\ref{subsec:pdf-intro-mesons} is applicable to any confining model in 1+1D, while that of Sec.~\ref{subsec:pdf-intro-baryons} applies only to non-abelian theories. Further details of specific theories will enter our analysis in Secs.~\ref{sec:meson}--\ref{sec:baryon}. 
In the upcoming calculation, for the sake of simplicity, we will assume the fractional momenta in the light-cone frame take on discrete values and each momentum state is normalized to one. 
This can be done by putting the system in a box of size L so that all momenta are quantized in units of 1/L. 
This helps us look past cumbersome normalization factors as well. 
We will go back to the continuous limit ($\mathrm{L} \rightarrow \infty$) when we compare against existing numerical results in the literature.

\subsection{Meson wavefunction}
\label{subsec:pdf-intro-mesons}

In keeping with our goal to make connection with the PDF, we work in the Infinite Momentum Frame (IMF) where the bound state is boosted to a very large light-cone momentum ($P^-$) and small $P^+$ component such that $P^+P^- = M_{\text{hadron}}^2$. 
The state of partons can then be expressed in terms of the fraction of the large momentum component $P^-$.
Assuming discrete values for this fractional momentum, 
the general quark--anti-quark state wavefunction can be written as 
\bea
|\psi\rangle = \sum_{i,\bar j} p_{i,\bar j}\delta_{i+\bar j,1}|i,\bar j \rangle  ,
\eea
where possible values of the discrete fractional light-cone momentum carried by the quark and the anti-quark, respectively, are denoted by $i$ and $\bar{j}$, and the $\delta$ function enforces momentum conservation.  
This is essentially the two-parton subsystem referred to in our conjecture.
For non-abelian gauge theories, the quark and anti-quark appear in a color anti-symmetric state, which suggests their momentum space wavefunction is completely symmetric. 
Hence, we suppress any explicit color indices in our calculation. 
The quark or the anti-quark reduced density matrix can be calculated as
\begin{align}
\rho_1 &= \sum_i \sum_{i'} \sum_{\bar j} p_{i,\bar j}p^*_{i',\bar j}\delta_{i+\bar j,1}\delta_{i'+\bar j,1}|i\rangle \langle i'|\nn\\
&= \sum_i |p_{i,1-i}|^2|i\rangle \langle i|.
\label{eq:rho1meson}
\end{align}
It is clear that the reduced density matrix is diagonal in this basis. 

As described in the previous section, the function that is actually extracted from any experiment is the PDF, which can be related to wavefunction of the hadron. 
To see that, note that the PDF of a fermion $q$ in the hadron $\mathcal{H}$ (with momentum $P^\mu$) is defined as
\begin{equation}
    f_q (x) = \int_{-\infty}^{\infty} \frac{dt}{2\pi} e^{-i t x (n \cdot P)}   \Big\langle \mathcal{H} \Big|    \mathcal{O}_2    \Big| \mathcal{H} \Big\rangle,
    \label{eq:pdfdef}
\end{equation}
where $x$ is the fractional momentum of the hadron carried by the parton $q$, $\mathcal{O}_2$ is a twist two operator
\begin{equation}
    \mathcal{O}_2= \frac{1}{2} \bar{\psi}_q (t n^\mu) \slashed{n} W_n (t n^\mu,0) \psi_q (0),
    \label{eq:twist2def}
\end{equation}
with $n^\mu$ being a lightlike vector whose spatial component is in the opposite direction of the hadron's spatial momentum vector, $W_n (a,b)$ is a Wilson line of the confining gauge group stretched between points $a$ and $b$, and $\psi(x)$ is the quark operator. 
In 1+1D, as eluded to earlier, the radiative corrections are power suppressed so that the Wilson line and any insertions from the Lagrangian drop out. The operator $\mathcal{O}_2$ is then just the quark number density operator and the PDF simply counts the average number density of quarks carrying a fraction $x$ of the hadron momentum. 
Combined with Eq.~\eqref{eq:rho1meson} for the case of the meson in 1+1D, $|p_{i,1-i}|^2$ is the discretized quark PDF with $x=i$, i.e.
\begin{equation}
    f_q(i) = |p_{i,1-i}|^2.
    \label{eq:defpdfmeson}
\end{equation}
In the continuous limit, we simply replace the discrete fractional momentum $i$ with the continuous variable $x$. Correct quark sum rule follows from $\Tr [\rho_1]=1$. 

$|p_{i,1-i}|^2$ must also obey the (discretized) momentum sum rule for the PDF, namely 
\bea
\sum_i i|p_{i,1-i}|^2 + \sum_{\bar j} {\bar j}|p_{1-\bar{j},\bar {j}}|^2 =1 .
\eea
By symmetry we infer that on average, the quark and anti-quark carry 1/2 of the hadron momentum so that 
\bea 
\sum_i i|p_{i,1-i}|^2 =\frac{1}{2}.
\label{eq:mesonconst}
\eea
This is automatically guaranteed if the PDF $|p_{i,1-i}|^2$ is symmetric under $i \leftrightarrow 1-i$, which is a known property of meson wavefunctions in 1+1D models.

We will use the von Neumann entropy of the reduced density matrix from Eq.~\eqref{eq:rho1meson}, namely
\bea 
S = - \sum_i |p_{i,1-i}|^2 \ln |p_{i,1-i}|^2 ,
\label{eq:defS}
\eea
and the kinetic energy of free quarks \cite{tHooft:1974pnl}, 
\bea
E =  \frac{m_q^2}{P^-}\sum_i |p_{i,1-i}|^2 \left(\frac{1}{i}+\frac{1}{1-i}\right),
\label{eq:kineticmeson}
\eea
in Eq.~\eqref{eq:defF} for mesons and minimize the resulting free energy. In deriving Eq.~\eqref{eq:kineticmeson} we use the fact that in the IMF, $P^+$ operator is identical to the Hamiltonian \cite{Brodsky:1997de}. 
Note should be taken that since the parton reduced density matrix of Eq.~\eqref{eq:rho1meson} is diagonal, the von Neumann and Shannon entropies of $|p_{i,1-i}|^2$ distribution are identical. From minimizing the free energy, we find 
\bea
|p_{i,1-i}|^2 \propto \exp \left[- \frac{m_q^2}{P^-T} \left(\frac{1}{i}+\frac{1}{1-i}\right)\right],
\eea
as would be expected for a thermal ensemble. For simplicity, we define
\begin{equation}
    \mathcal{T}^2 \equiv P^- T.
    \label{eq:curlyTdef}
\end{equation}
Since $P^-$ and $T$ appear in our calculation always in this combination, we work with $\mathcal{T}$ for the rest of our calculation. 
Taking the continuum limit and normalizing the distribution, we find 
\bea
f_q(x)=|p(x,1-x)|^2 =  \frac{\exp\left[-\frac{m_q^2}{\mathcal{T}^2}\left(\frac{1}{x}+\frac{1}{1-x}\right)\right]}{{\displaystyle \int_0^1 dx} \exp \left[-\frac{m_q^2}{\mathcal{T}^2}\left(\frac{1}{x}+\frac{1}{1-x}\right)\right]}.
\label{eq:Ansatz}
\eea
The value of $\mathcal{T}$ is still to be determined and this is where the details of the specific model will come into play. This will be determined by minimizing the expectation value of the full hadron Hamiltonian with this thermal ansatz. 
In the upcoming section we use this ansatz to predict mass and PDF of mesons in various models in 1+1D.

For massless partons ($m_q=0$), Eq.~\eqref{eq:Ansatz} suggests $f_q(x)$ is uniform in $x$, agreeing with existing results in the literature \cite{tHooft:1974pnl}.
This PDF maximizes the entropy of Eq.~\eqref{eq:defS}, implying the meson wavefunction maximizes the entanglement entropy of a single parton, inside this two-parton system, in the massless quark limit.


\subsection{Baryon wavefunction}
\label{subsec:pdf-intro-baryons}

Let us study our conjecture for baryonic systems in $N=3$ theories. (The case of larger gauge groups are addressed at the end of this subsection.) These baryons include three identical quarks in a color anti-symmetric configuration. 
The state of the system (ignoring the color indices) in discretized momentum space is written as 
\bea
 |\psi \rangle = \sum_{ijk=0}^{1} p_{i,j,k} \delta_{i+j+k,1}|i j k\rangle .
 \label{eq:Ba1}
\eea
The associated density matrix is
\bea
\rho = |\psi \rangle \langle \psi | =  \sum_{ijk=0}^{1}\sum_{\bar i \bar j \bar k=0}^{1} p_{i,j,k} p^*_{\bar i,\bar j,\bar k}\delta_{i+j+k,1} \delta_{\bar i+\bar j+\bar k,1}|i j k\rangle \langle \bar i \bar j \bar k|.
\label{eq:2}
\eea

Now we trace over one of the quarks, say the one with momentum fraction k,
\bea
\text{Tr}_k\Big[\rho \Big] \equiv  \bar \rho= \sum_l \sum_{ij}^{1-l}\sum_{\bar i \bar j }^{1-l} p_{i,j,l} p^*_{\bar i,\bar j,l}\delta_{i+j+l,1} \delta_{\bar i+\bar j+l,1}|i j \rangle \langle \bar i \bar j |,
\label{eq:Barho2}
\eea
where $l$ is the fraction of momentum carried by the traced out quark, and we have used momentum conservation to put the $1-l$ upper bound on the sums. 
We can now rewrite the expression above as a weighted sum over  two-quark reduced density matrices, carrying a total momentum fraction $1-l$
\bea
 \bar \rho =   \sum_l \mathcal{N}_l  \rho_{2,l},
 \label{eq:BarhobarNlfactor}
\eea
with
\begin{eqnarray}
\label{eq:defrho2}
\rho_{2,l} &\equiv  & \sum_{ij=0}^{1-l}\sum_{\bar i \bar j =0}^{1-l} {\bf p}_{i,j,l}{\bf  p}^{*}_{\bar i,\bar j,l}\delta_{i+j,1-l} \delta_{\bar i+\bar j,1-l}|i j \rangle \langle \bar i \bar j |, \\
\label{eq:defNl}
\mathcal{N}_l  &\equiv & \sum_{i,j =0}^{1-l} |p_{i,j,l}|^2 \delta_{i+j,1-l} , \\
\label{eq:defpbold}
{\bf p} _{i,j,l} & \equiv & \frac{p_{i,j,l}}{\sqrt{\mathcal{N}_l}}.
\end{eqnarray}
$\rho_{2,l}$ is now an $l$-dependent two-quark reduced density matrix. This density matrix corresponds to a pure state associated with a \textit{fixed-momentum two-parton subsystem} of the baryon 
\bea
| \psi_l^{(2)} \rangle =  \sum_{ij=0}^{1-l} {\bf p}_{i,j,l}\delta_{i+j,1-l}|i j \rangle ,
\label{eq:2quark}
\eea
so that the problem is now the same as the meson system repeated for each value of $l$. 

We apply our free energy ansatz for $every$ two-parton subsystem ($\rho_{2,l}$) independently in the weighted sum in Eq.~\eqref{eq:BarhobarNlfactor}. 
As before, we can now trace over the second quark in every $\rho_{2,l}$ to get the one quark density matrix
\bea
\rho_{1,l} = \sum_{i}  |{\bf p}_{i,1-i-l,l}|^2   | i\rangle \langle i|.
\label{eq:rho1Ba}
\eea
Note that $\Tr [\rho_{1,l}]=1$ and it is a well-defined reduced density matrix. 
We stress that $\rho_{1,l}$ is not the reduced density matrix of a single quark if we had traced out the other two quarks in the full baryon density matrix,
rather it can be considered a \textit{fixed-momentum} single quark reduced density matrix within a two-quark subsystem of the hadron.

The von Neumann entropy of this fixed-momentum density matrix is given by 
\begin{equation}
    S_l = - \sum_i |{\bf p}_{i,1-i-l,l}|^2 \ln |{\bf p}_{i,1-i,l}|^2.
    \label{eq:defSlBa}
\end{equation}
Since the two-quark subsystem carried the total momentum fraction $1-l$, their free parton kinetic energy in the light-cone frame can be written as 
\bea
 E_l =  \frac{m_q^2}{P^-} \sum_{i=0}^{1-l}  |{\bf p}_{i,1-i-l,l}|^2 \left(  \frac{1}{i} + \frac{1}{1-l-i} \right),
 \label{eq:defKlBa}
\eea
where $m_q$ is again denoting the bare quark mass and $P^-$ is the light-cone momentum. 
We use Eqs.~\eqref{eq:defSlBa}--\eqref{eq:defKlBa} in the free energy ansatz of our conjecture in Eq.~\eqref{eq:defF}. 

In other words, we conjecture the baryon wavefunction minimizes free energy of \textit{every fixed-momentum two-parton subsystems}. 
An intuitive way to justify using this particular density matrix is that we have based our free energy arguments on the nature of interactions between quarks. 
In particular, it is the interaction between $two$ quarks that is local and, hence, would lead to maximum entanglement between them in the momentum space.

Minimizing this free energy, we find
\bea
 |{\bf p}_{i,1-i-l,l}|^2  = \frac{ \exp \left[ -\frac{m_q^2}{\mathcal{T}^2} \left(  \frac{1}{i} + \frac{1}{1-l-i}\right) \right]}{\sum_{i=0}^{1-l}  \exp \left[ -\frac{m_q^2}{\mathcal{T}^2} \left(  \frac{1}{i} + \frac{1}{1-l-i}\right) \right]},
 \label{eq:pboldexpr}
\eea
where $\mathcal{T}$ is defined in Eq.~\eqref{eq:curlyTdef}. 
Combining this with Eqs.~\eqref{eq:defNl}--\eqref{eq:defpbold}, we find
\bea
\frac{|p_{i,1-i-l,l}|^2}{\sum_{i=0}^{1-l} |p_{i,1-l-i,l}|^2} =  \frac{ \exp \left[ -\frac{m_q^2}{\mathcal{T}^2} \left(  \frac{1}{i} + \frac{1}{1-l-i}\right) \right]}{\sum_{i=0}^{1-l}  \exp \left[ -\frac{m_q^2}{\mathcal{T}^2} \left(  \frac{1}{i} + \frac{1}{1-l-i}\right) \right]},
\label{eq:pBaexpr}
\eea
from which we can write
\bea
|p_{i,1-l-i,l}|^2 = \bar{\mathcal{A}}(l)   \exp \left[ -\frac{m_q^2}{\mathcal{T}^2} \left(  \frac{1}{i} + \frac{1}{1-l-i}\right) \right]. 
\eea
Since the quarks are identical, the normalization factor $\bar{\mathcal{A}}(l)$ should take a form such that
the result is manifestly symmetric under interchange of any two of them. Thus, we should find
\bea
 |p_{i,1-l-i,l}|^2 = \mathcal{A}^{-1}   \exp \left[ -\frac{m_q^2}{\mathcal{T}^2} \left(  \frac{1}{i} + \frac{1}{1-l-i}+ \frac{1}{l}\right) \right],
 \label{eq:pBaansatz}
\eea
where now the overall normalization factor is given by
\bea 
\mathcal{A} =  \sum_{l=0}^1 \sum_{i=0}^{1-l}  \exp \left[ -\frac{m_q^2}{\mathcal{T}^2} \left(  \frac{1}{i} + \frac{1}{1-l-i}+ \frac{1}{l}\right) \right].
\label{eq:NBanorm}
\eea

Let us emphasize again that the fixed-momentum single parton density matrix we used is not the same as the reduced density of a single quark in the baryon. The latter can be derived from Eq.~\eqref{eq:Ba1} as\footnote{$|p_{i,l,1-i-l}|^2$ can be thought as a joint probability distribution for two random variables $i$ and $l$. 
It is worth mentioning that the PDF ansatz in Eq.~\eqref{eq:pBaansatz} could be derived from a free energy function made of the free parton kinetic energy of three quarks and the Shannon entropy of this joint probability distribution. However, we believe the physical interpretation of \textit{fixed-momentum} two-parton subsystem lends itself better to an intuitive justification of minimum free energy principle. }
\begin{equation}
    \rho_1 = \sum_i\sum_{l}^{1-i} |p_{i,l,1-i-l}|^2 |i \rangle \langle i |.
    \label{eq:rho1baryon}
\end{equation}

Similar to the meson case in the previous section, we can show that in these toy models in 1+1D, the PDF of a quark coincides with average number density of quarks carrying momentum $i$, i.e. using Eq.~\eqref{eq:rho1baryon} we have 
\begin{equation}
    f_q(i) = 3\sum_{l=0}^{1-i} |p_{i,l,1-i-l}|^2,
    \label{eq:defpdfbaryons}
\end{equation}
where the factor of three indicates that we have three identical quarks. 

So far we assumed the momentum can only take discrete values. Going to the continuous case is now straightforward. We denote the continuous version of $p_{i,j,k}$ as
\begin{equation}
    p_{i,j,l} \rightarrow \Phi (x,y,z),
    \label{eq:pPhicontinuous}
\end{equation}
where now the fractional momentum variables $(x,y,z)$ are all continuous and in the range $[0,1]$. The continuous momentum version of Eqs.~\eqref{eq:pBaansatz}--\eqref{eq:NBanorm} are now 
\bea
 |\Phi (x,1-z-x,z)|^2 = \mathcal{A}^{-1}  \exp \left[ -\frac{m_q^2}{\mathcal{T}^2} \left(  \frac{1}{x} + \frac{1}{1-z-x}+\frac{1}{z}\right) \right],
 \label{eq:BAnsatz}
\eea
where now the overall normalization factor is given by
\bea 
\mathcal{A}=  \int_0^1 dx \int_0^{1-x} dz  ~ \exp \left[ -\frac{m_q^2}{\mathcal{T}^2} \left(  \frac{1}{x} + \frac{1}{1-z-x}+\frac{1}{z}\right) \right].
\label{eq:BNorm}
\eea
Putting these equations back in Eq.~\eqref{eq:defpdfbaryons}, we find the final form of our ansatz for quarks' PDF in a baryon

\bea
f_q(x) = 3\frac{ {\displaystyle \int_0^{1-x} dz }~  \exp \left[ -\frac{m_q^2}{\mathcal{T}^2} \left(  \frac{1}{x} + \frac{1}{1-z-x}+\frac{1}{z}\right) \right]}{ {\displaystyle\int_0^1 dx \int_0^{1-x} dz } ~ \exp \left[ -\frac{m_q^2}{\mathcal{T}^2} \left(  \frac{1}{x} + \frac{1}{1-z-x}+\frac{1}{z}\right) \right]}.
\label{eq:ansatzbaryon}
\eea

Finally, we should note that our discussion in this section can be generalized to the case of $N \geqslant 4$ colors. 
To do this, we should first trace out a quark to get the \textit{fixed-momentum} $(N-1)$-quark subsystem; this process can be repeated $N-2$ times, as illustrated above, to reach the fixed-momentum two-parton subsystem.
We can then minimize every such subsystems' free energy and work our way back to the full wavefunction. The resulting PDF ansatz will look like
\begin{equation}
    f_q (x) = N \frac{ {\displaystyle \left( \prod_{i=1}^{N-2}  \int_0^{1-x-\sum_{j=1}^{i-1} z_j}  dz_i \right) } ~ \exp \left[ -\frac{m_q^2}{\mathcal{T}^2} \left(  \frac{1}{x} + \frac{1}{1-x-\sum_{k=1}^{N-2} z_k} + \sum_{k=1}^{N-2} \frac{1}{z_k}  \right)   \right] }{{\displaystyle \int_0^1  dx  \left( \prod_{i=1}^{N-2} \int_0^{1-x-\sum_{j=1}^{i-1} z_j}  dz_i \right) }~  \exp \left[ -\frac{m_q^2}{\mathcal{T}^2} \left(  \frac{1}{x} + \frac{1}{1-x-\sum_{k=1}^{N-2} z_k} + \sum_{k=1}^{N-2} \frac{1}{z_k}  \right)   \right] } . 
    \label{eq:ansatzNlarger}
\end{equation}

We should also note that in the massless quark limit the free partons kinetic energy vanishes and our conjecture becomes equivalent to maximizing the entanglement entropy of the fixed-momentum two-parton subsystem. From Eq.~\eqref{eq:ansatzNlarger} we find that in this limit
\begin{equation}
    f_q (x) = N (N-1) (1-x)^{N-2},
    \label{eq:pdfbaryonmq0}
\end{equation}
in agreement with the existing results in the literature \cite{Hornbostel:1988fb}. 
In the upcoming section we compare our ansatz in Eq.~\eqref{eq:Ansatz}, for mesons, and Eq.~\eqref{eq:ansatzbaryon}, for baryons, with existing results in the literature away from the $m_q=0$ limit.

\section{Meson Spectrum and PDF}
\label{sec:meson}

\subsection{ Schwinger Model}
\label{subsec:schwinger}
The Schwinger model \cite{Schwinger:1962tp} is the simplest example of an exactly solvable gauge theory. This is an abelian gauge theory and will be a stepping stone to non-abelian gauge theories.
We start with the action 
\begin{eqnarray}
\label{eq:SchAction}
S &= \int d^2x \left(\frac{1}{2}\mathcal{E}^2 +\bar \psi i\gamma^{\mu}(\partial_{\mu}-m_q-gA_{\mu})\psi\right) \\
& \equiv \int d^2x \left(\frac{1}{2}\mathcal{E}^2 +i\bar \psi \gamma^{\mu}\partial_{\mu}\psi- i m_q \bar \psi\psi -iA_{\mu}j^{\mu}\right), \nn
\end{eqnarray}
where $\mathcal{E}$ is the field strength of the abelian gauge group.
For the weakly interacting fermion theory, $m_q \gg g$, the spectrum can be obtained using perturbation theory~\cite{Coleman:1974bu,Coleman:1985rnk}. 
For the strongly interacting regime, $g \gg m_q$, the Schwinger model can be solved  by appealing to a duality with a theory of bosons \cite{Schwinger:1962tp}. 
After bosonization, we simply obtain the action of a massive scalar field with a normal ordered  cosine interaction term 
\bea
 S=  \frac{1}{8\pi}\int d^2x \left((\partial_{\mu}\phi)^2-\frac{g^2}{\pi}\phi^2 +\frac{m}{g}:\cos \phi :\right),
\eea
where $\phi (x)$ is the meson wavefunction in momentum space and the PDF is $f_q(x)= |\phi (x)|^2$.
The intermediate regime $m_q \sim g$, however, still needs to be solved numerically. 

In Eq.~\eqref{eq:SchAction}, the photon can be eliminated using gauge redundancy and equations of motion, i.e. it is not a propagating mode in 1+1D.  
Keep in mind that we work in the infinite momentum frame with $P^- \rightarrow \infty$, where $P^+$ acts as the Hamiltonian. 
In this frame the right-handed component of the fermion field, $\psi_R$, is not a propagating degree of freedom either and can be eliminated using equations of motion. 
Having integrated out these fields, we find an effective four-fermion interaction term in the Hamiltonian written in position space as \cite{Coleman:1974bu,Coleman:1985rnk} 
\bea
H_{\text{int}}(x^0) = g^2\int dx^1 dy^1 J(x^0,x^1)|x^1-y^1|J(x^0,y^1),
\eea
where $J$ is the bilinear left handed fermion current $ \psi_L^{\dagger} \psi_L$. We see that the potential increases linearly with distance, manifesting the confinement.   
Given a two-parton meson state with a wavefunction $\phi(x,1-x)$, where $x$ is the momentum fraction of $P^-$ carried by the quark, one way of solving for the spectrum is to look at the effective light-cone Schr\"{o}dinger equation obeyed by the two-parton wavefunction \cite{Coleman:1985rnk,Hornbostel:1988fb} 
\small
\bea 
M^2_{\text{hadron}}\phi(x)=m_q^2\left(\frac{1}{x}+\frac{1}{1-x}\right)\phi(x)-\frac{g^2}{2\pi}\text{P}\int dy \frac{\phi(y)-\phi(x)}{(y-x)^2}+\frac{g^2}{\pi}\int_0^1dy \phi(y),
\label{eq:Schroedinger}
\eea
\normalsize
where P indicates the principal value of the integral. 
However, as per our philosophy, instead of solving this complicated equation, we want to see if a simple variational principle can give us the correct answer. To do that, we need the expectation value of the Hamiltonian ($P^+$)
\begin{align}
\label{eq:Hsch}
& M_{\mathrm{hadron}}^2 =\langle \phi| P^{-} P^+ |\phi \rangle =\langle \phi| P^{-} (H^0+H_{\text{int}}) |\phi \rangle \\
& =m_q^2\int dx |\phi(x)|^2\left(\frac{1}{x}+\frac{1}{1-x}\right)+ \frac{g^2}{2\pi}\int dx dy \frac{|\phi(x)-\phi(y)|^2}{(x-y)^2}+\frac{g^2}{\pi}\int dx dy \phi(x) \phi^*(y). \nn
\end{align}
\normalsize
Here $M_{\mathrm{hadron}}^2$ is the invariant mass squared of the meson bound state. 

We now plug in the ansatz from Eq.~\eqref{eq:Ansatz} and minimize $M_{\mathrm{hadron}}^2$ over $\mathcal{T}$ to find the mass and wavefunction of the meson. We assume the wavefunction is real, or has a phase factor that is independent of the momentum fraction $x$. 
Note that this is simply a variational method for calculating the meson wavefunction and mass, where the wavefunction ansatz minimizes partons free energy defined in Eq.~\eqref{eq:defF}. 

\begin{figure}
\centering
\includegraphics[width=\linewidth]{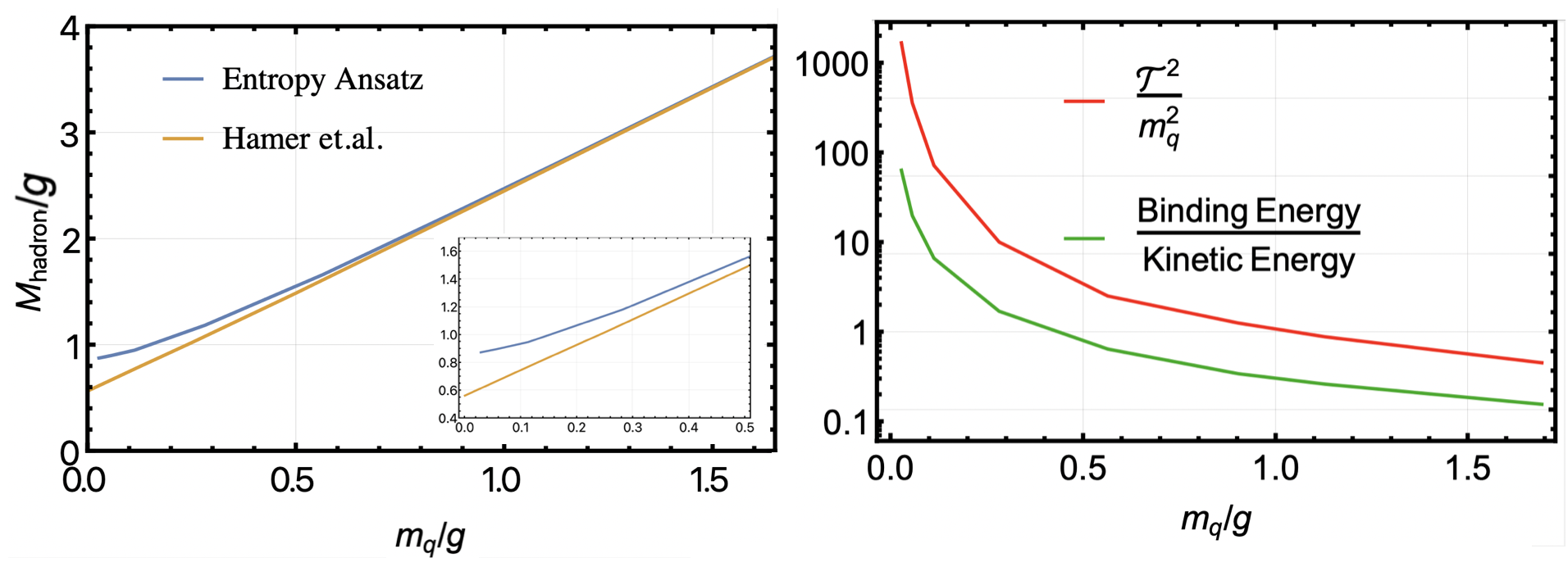}
\caption{\textbf{Left:} Mass spectrum of the lowest meson bound state as a function of parton mass for the Schwinger model. The blue (amber) line shows our prediction from minimizing the free energy of the hadron (existing numerical calculation from solving the time-independent light-cone Schr\"{o}dinger equation by Hamer et al.~\cite{Sriganesh:1999ws}. 
We find perfect agreement between the two methods in the limit of $m_q \gg g$, corroborating our conjecture in this limit. 
\textbf{Right:} Effective temperature ($\mathcal{T}^2/m_q^2$) (blue) and the ratio of binding energy to kinetic energy (yellow) for the lowest meson bound states. We find that at $m_q \gg g$ limit, the system asymptote to a zero temperature gas. }
\label{fig:Sch}
\end{figure}  

Results of this calculation, and comparison with existing numerical results in the literature \cite{Sriganesh:1999ws} are shown in Fig.~\ref{fig:Sch}. 
The left panel shows the bound state spectrum for the lowest state as a function of the parton mass $m_q$. 
For large masses ($m_q/g \gsim 1)$ we find perfect agreement between the lattice result and our prediction. This can be attributed to the fact that this is the weakly coupled regime and the theory behaves like a pair of nearly free quark and anti-quark, which is essentially a system at zero effective temperature.
In this regime, we expect the system to be very weakly entangled and the state is just a product state of quark and anti-quark each carrying half of the meson momentum. 

As we move towards smaller masses, the agreement with lattice worsens, although even at masses as low as $m_q/g= 0.25$ the error is only at $\sim 10 \% $. 
The deviation from the lattice result can be traced to the behavior of our ansatz near the extremes $x\sim 0,1$. At $x \sim 0$, the PDF goes as $e^{- \alpha/x}$, which drops more abruptly than suggested by the Schr\"{o}dinger equation from Eq.~\eqref{eq:Schroedinger} ($\sim x^{\alpha}$) \cite{tHooft:1974pnl} (see also Ref.~\cite{Anand:2021qnd} for recent progress on this). 
As we move towards $m_q=0$, the derivative of the ansatz approaches a delta function near $x \sim 0, 1$;  consequently, the potential energy, which is sensitive to the derivative of the wavefunction, asymptotes to a constant instead of dwindling to zero. 
This also leads to a discontinuity at the point $m_q=0$. Nonetheless, it is clear that setting $m_q=0$ exactly in Eq.~\eqref{eq:Ansatz} predicts a uniform wavefunction, in agreement with existing results~\cite{Schwinger:1962tp}. 
This suggests that the system is not well simulated as a thermal ensemble for finite but small $m_q/g$ values. In the final section, we will discuss some plausible ways in which our variational principle could be modified to have better agreement.

In the right panel of Fig.~\ref{fig:Sch} we show the ratio of binding energy to the kinetic term. We see non-negligible contribution from the potential term to the Hamiltonian of the hadron. Note that this potential is not included in our free energy term. 
We also show a plot of the exponent $\mathcal{T}^2/m_q^2$ in our PDF anstaz, which we see is monotonically decreasing as expected since in this limit the interaction between partons is becoming weaker. 
This is a more meaningful quantity to consider as against $\mathcal{T}$, since we are trying to compare the temperature of different systems (in terms of the mass of their constituents).

\begin{figure}
\centering
\includegraphics[width=\linewidth]{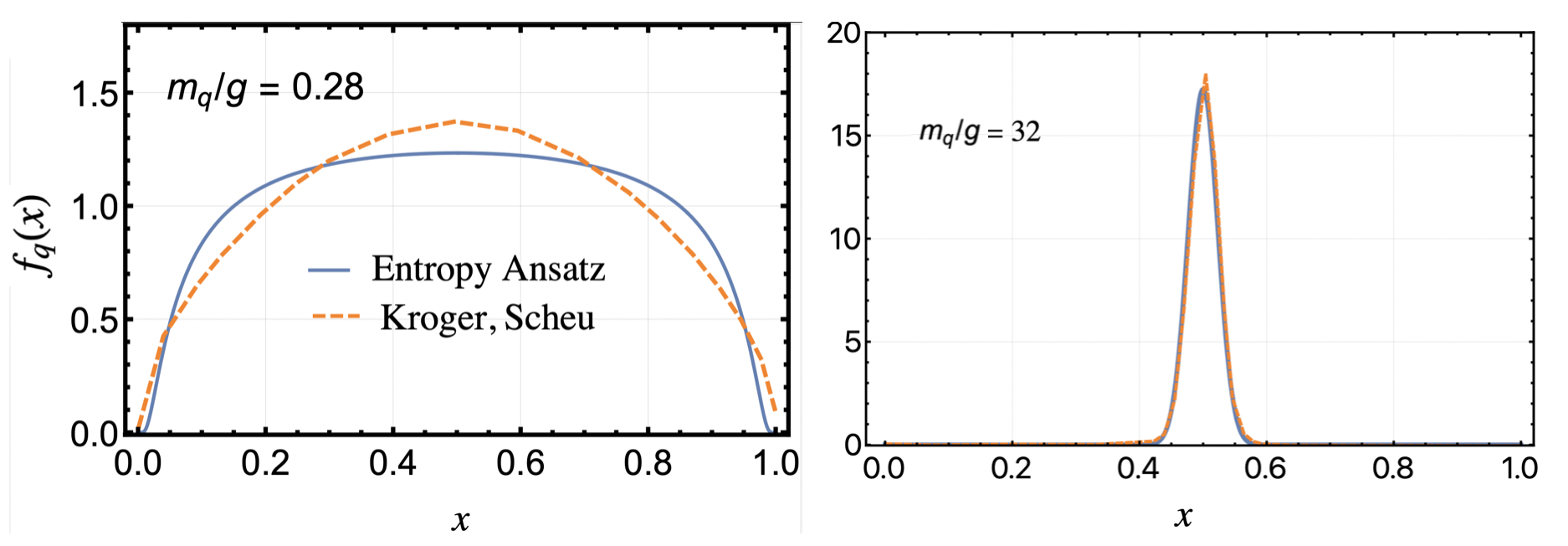}
\caption{ The PDF of the Schwinger model ground state meson for $m_q/g=0.28$ (\textbf{left}) and $m_q/g=32$ (\textbf{right}). The prediction from our minium free energy principle (existing results from solving the time-independent like-cone Schr\"{o}dinger equation taken from work of Kroger et al.~\cite{Kroger:1998se}) is shown in blue (orange). We find perfect agreement in the heavy parton mass limit, while even at lower masses our ansatz (Eq.~\eqref{eq:Ansatz}) captures the general behavior of the PDF.}
\label{fig:SPDF}
\end{figure}  

We also show PDFs for different values of $m_q$ in Fig.~\ref{fig:SPDF}. While in large $m_q$ limit we find perfect agreement with existing results \cite{Kroger:1998se},
at low quark masses the agreement starts disappearing, even though the general shape of the PDF is still captured by our calculation. 
In the limit of exact $m_q=0$, our results again agree with the conventional calculation. 

While the kinetic term is shared between the free energy (Eq.~\eqref{eq:defF}) and the Hamiltonian (Eq.~\eqref{eq:Hsch}), the second term in each of these quantities (respectively the entanglement entropy or the potential term) are ostensibly unrelated objects that do not know anything about each other. 
Hence, our free energy and the Hamiltonian are distinct quantities and the fact that our ansatz reproduces both the mass spectrum and the wavefunction is highly non-trivial. 

\subsection{'t Hooft model}
\label{subsec:thooft-meson}
The extension of the Schwinger model to the non-abelian case with large number of colors is the the 't Hooft model \cite{tHooft:1974pnl}. 
The action for the model is
\bea
S= \int d^2x\left(-\frac{1}{4}F_{\mu \nu}^aF^{\mu \nu a} +\bar \psi (i \slashed{D}-m_q) \psi \right),
\label{eq:thooftS}
\eea
with $\psi$ in the fundamental representation of the gauge group SU($N$). 
The story for mesons is similar as for the Schwinger model, in that we still have a quark--anti-quark state but now in a linear combination of colors. 
 It can be shown that the expectation value of the Hamiltonian $P^+P^-$ in a two-parton meson state can be written as \cite{Hornbostel:1988fb}
\bea
M_{\mathrm{hadron}}^2=m_q^2\int dx |\phi(x)|^2\left(\frac{1}{x}+\frac{1}{1-x}\right)+\frac{g^2}{4\pi}\frac{N^2-1}{N}\int dx \int dy \frac{|\phi(x)-\phi(y)|^2}{(x-y)^2},
\label{eq:tHooftH}
\eea
where $\phi (x)$ is the meson wavefunction in momentum space and $f_q(x)= |\phi (x)|^2$.

Similar to the Schwinger model, we can now put our ansatz for the meson wavefunction\footnote{Again we assume the wavefunction is real or has no $x$ dependent phase factor.} (Eq.~\eqref{eq:Ansatz}) in Eq.~\eqref{eq:tHooftH} to minimize the expectation value of the Hamiltonian and thereby calculate the intrinsic $\mathcal{T}$ and, thus, the wavefunction and the meson mass. 
For $m_q=0$, our ansatz in Eq.~\eqref{eq:Ansatz} predicts a uniform PDF and $M_{\mathrm{hadron}}=0$, in agreement with results in the literature from solving Eq.~\eqref{eq:tHooftH}. We repeat this calculation for different $m_q$ values to calculate the meson mass and PDF. 

\begin{figure}
    \centering
    \resizebox{\columnwidth}{!}{
    \includegraphics{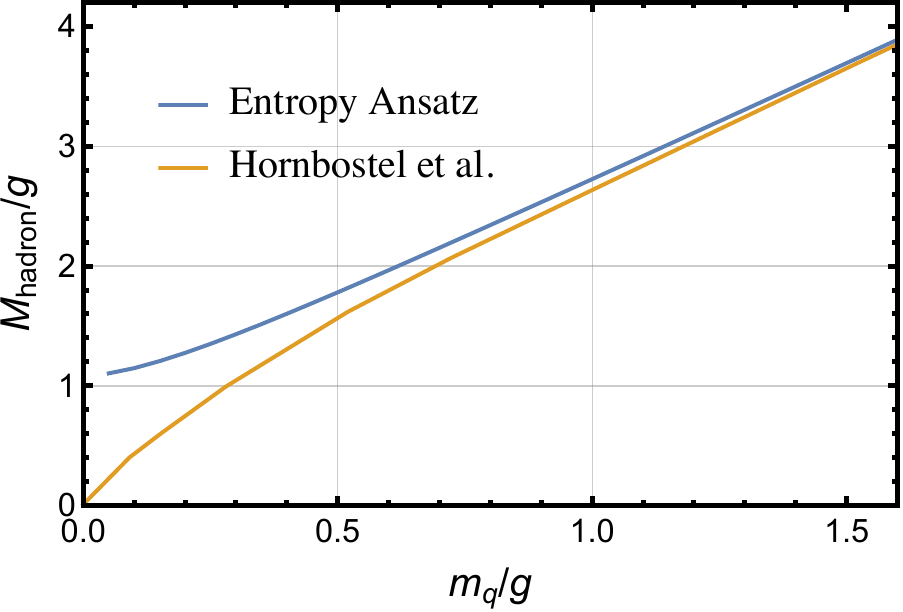}
    \includegraphics[scale=1.07]{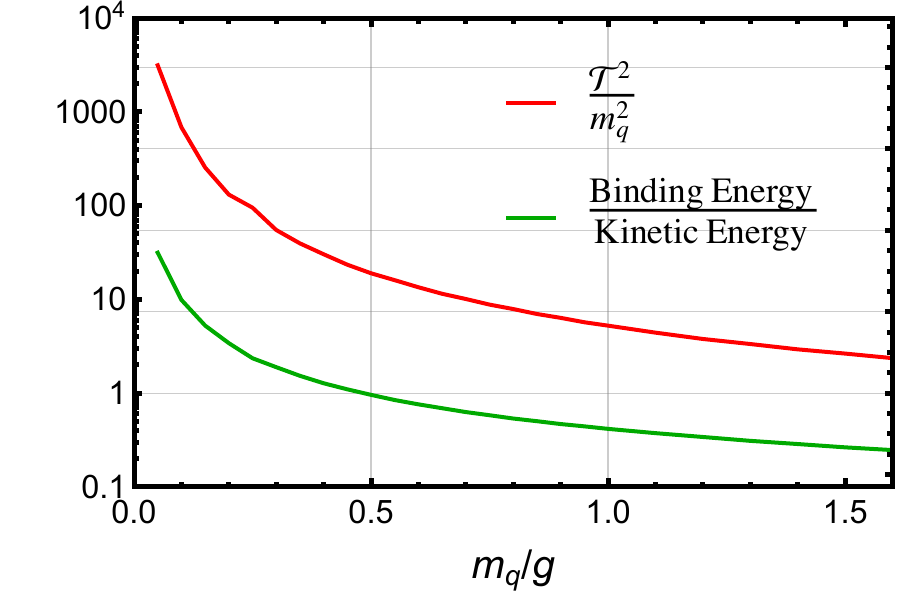}
    }
    \caption{\textbf{Left:} Ground state meson's mass for $N$=3 derived from our calculation (blue), compared to the existing results from Hornbostel et al. \cite{Hornbostel:1988fb} (amber). We find a perfect agreement in the heavy quark limit, while for small but finite quark masses the agreement is not as well.  \textbf{Right:} The effective intrinsic temperature of the meson (see Eq.~\eqref{eq:curlyTdef}) and the contribution of hadron binding energy to its mass. We find that at $m_q \gg g$ limit, the system asymptote to a zero temperature gas.   }
    \label{fig:Mesonmass}
\end{figure}
\begin{figure}
    \centering
    \resizebox{0.7\columnwidth}{!}{
    \includegraphics{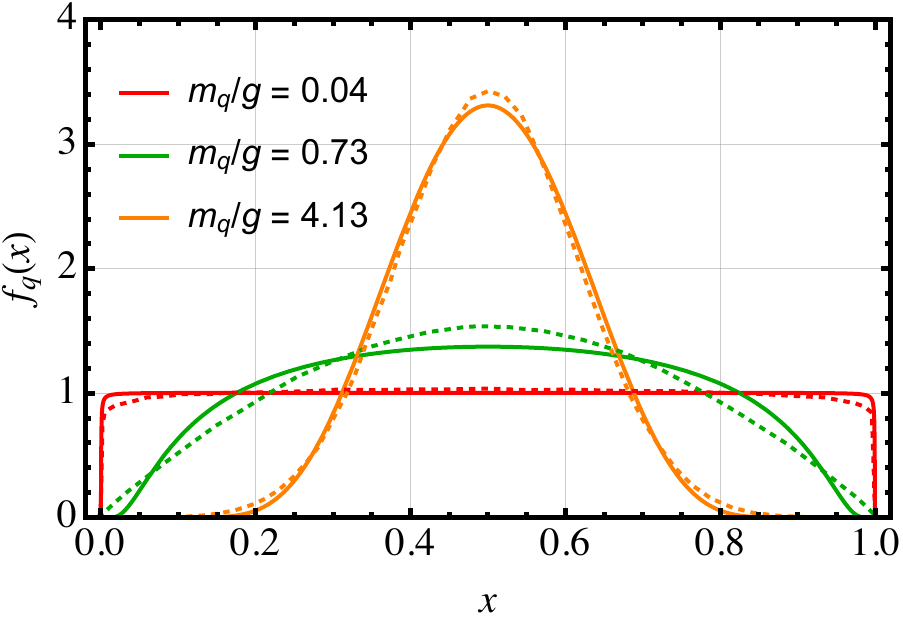}
    }
    \caption{Mesons' PDFs for $N$=3 for different values of $m_q/g$. The solid (dotted) lines denote the result of our calculation from minimizing the free energy of the meson (numerical results of Ref.~\cite{Hornbostel:1988fb} from solving a time-independent light-cone Schr\"{o}dinger equation).
    We find particularly good agreement in the limit of $m_q/g \gg 1$ (heavy quarks limit) while even at lower quark masses our ansatz (Eq.~\eqref{eq:Ansatz}) captures the general behavior of the existing results in the literature. }
    \label{fig:Mesonpdf}
\end{figure}

Figure \ref{fig:Mesonmass} shows the comparison of the ground state meson mass spectrum (as a function of the parton mass for $N=3$) with a numerical computation \cite{Hornbostel:1988fb}. Similar to the Schwinger model, the result is exact in the limit of $m_q/g \gg 1$ and worsens at lower masses. 
The right panel of Fig.~\ref{fig:Mesonmass} shows non-negligible contribution from the potential term to the hadron Hamiltonian for values of $m_q/g$ that we find good agreement with literature. This re-emphasizes the fact that our alternative principle, which has no knowledge of the potential in calculating the wavefunction ansatz, is truly a new approach to calculating properties of hadrons in this theory. 
At finite but small $m_q$ values our prediction does not follow the existing numerical results, signaling the need for improvement in our method, which we will briefly comment on in the conclusion.

Figure~\ref{fig:Mesonpdf} shows a comparison of the meson PDF for $N=3$ with existing numerical computation \cite{Jia:2017uul}. We again find that our ansatz captures the general feature of the PDF for different values of $m_q$; it particularly works well in the limit of $m_q \gg g$.

\section{Baryon Spectrum and PDF}
\label{sec:baryon}

For non-abelian gauge groups we also have baryonic states in the spectrum. For an SU($N$) theory, this is a state of $N$ identical quarks in anti-symmetric color configuration, so their momentum space wavefunction is symmetric. 
For simplicity, we focus on the case of $N=3$ in our study of the baryons.

As explained before, we decompose the total Hilbert space into the Fock space of three individual quarks.
For 1+1D, for the lightest baryonic states, the contribution from higher Fock states is considerably suppressed \cite{Hornbostel:1988fb}.
Let us assume one of the quarks carries momentum fraction $x$ of the whole hadron, while another one carries the total momentum fraction $z$; this leaves momentum fraction $1-x-z$ for the third quark.

For this state, we can once again compute the expectation value of the Hamiltonian ($P^+P^-$) given the action Eq.~\eqref{eq:thooftS},

\small
\begin{align}
\label{eq:Hbaryon}
 M_{\mathrm{hadron}}^2 &=m_q^2 \int_0^1 dz \int_0^{1-z} dx \left(\frac{1}{x}+\frac{1}{1-z-x}+\frac{1}{z}\right) |\Phi(x,1-x-z,z)|^2 \\
 &+ \frac{g^2}{2\pi}\frac{3}{2}\frac{3^2-1}{3} \int_0^1 dz\int_0^{1-z} dx_1 \int_0^{1-z} dx_2 \frac{|\Phi(x_1,1-z-x_1,z) -\Phi(x_2,1-z-x_2,z)|^2}{(x_1-x_2)^2}, \nn
\end{align}
\normalsize
where factors of $3$ refer to number of colors. 
The expectation value of the Hamiltonian has the same form as that for a meson, now generalized to the baryonic state. In particular, since we restrict ourselves to a Fock space of 3 quarks, the binding energy is just the sum over the binding energies for three quark pairs.
We can now use the ansatz of Eq.~\eqref{eq:BAnsatz} with the normalization from Eq.~\eqref{eq:BNorm} in Eq.~\eqref{eq:Hbaryon} and find minimum $M_{\mathrm{hadron}}$ value, i.e. the ground state baryon mass, and the intrinsic temperature $\mathcal{T}$ that minimizes the Hamiltonian. 
Putting this $\mathcal{T}$ back in Eq.~\eqref{eq:ansatzbaryon}, we find the final form of the PDF for this state.  

For the massless quark case, i.e. $m_q=0$, we find a uniform distribution for the wavefunction $\Phi$,
\begin{equation}
    |\Phi(x,1-x-z,z)| = \sqrt{2}.
    \label{eq:mq0pdfbar}
\end{equation}
From this we also find 
\begin{equation}
    f_q (x) = 3\int_0^{1-x} dz |\Phi(x,1-x-z,z)|^2 =6(1-x),
    \label{eq:pdfbaryondef}
\end{equation}
and $M_{\mathrm{hadron}}=0$, in agreement with the existing results in the literature \cite{Hornbostel:1988fb}. 

\begin{figure}
    \centering
    \resizebox{\columnwidth}{!}{
    \includegraphics[scale=1]{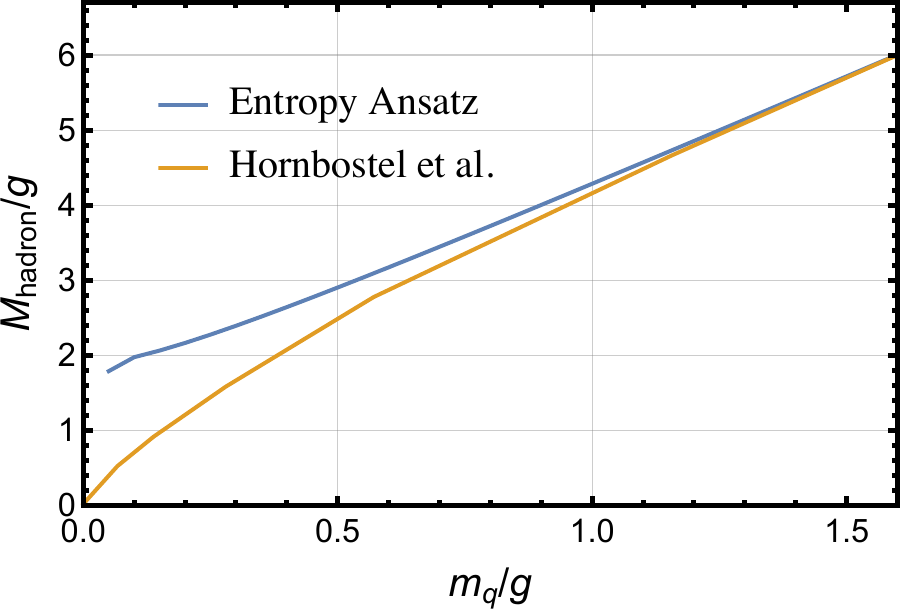}
    \includegraphics[scale=1.07]{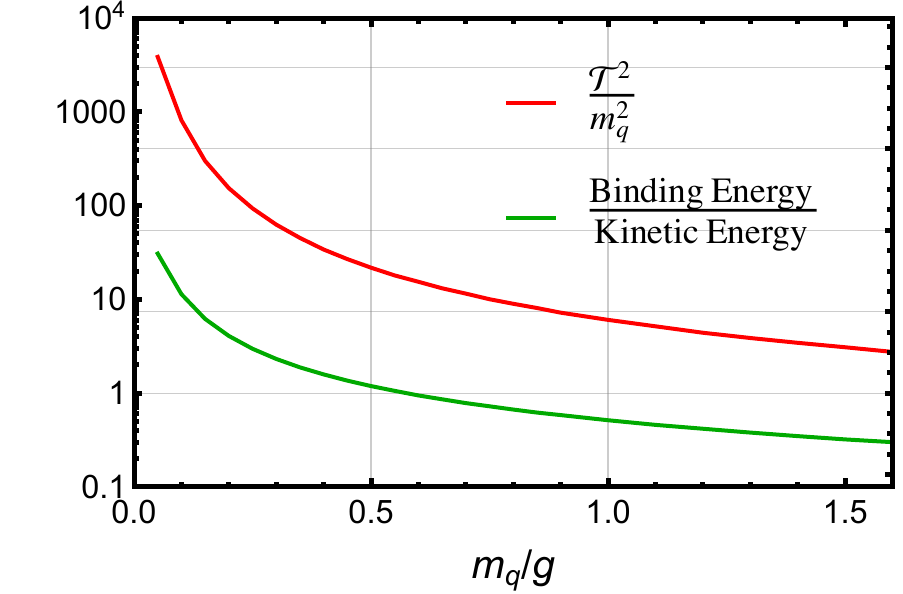}
    }
    \caption{\textbf{Left:} Ground state baryon's mass for $N$=3 derived from our calculation (blue), compared to the existing results in the literature (orange) \cite{Hornbostel:1988fb}. We find a great agreement in the heavy quark limit and for $m_q=0$, while for small but finite quark masses the agreement is not as great (we speculate about the potential culprits in the conclusion). \textbf{Right:} The effective intrinsic temperature of the baryon (see Eq.~\eqref{eq:curlyTdef}) and the contribution of hadron binding energy to its mass. We find that at $m_q \gg g$ limit, the system asymptote to a zero temperature gas.   }
    \label{fig:Baryonmass}
\end{figure}
\begin{figure}
    \centering
    \resizebox{0.7\columnwidth}{!}{
    \includegraphics{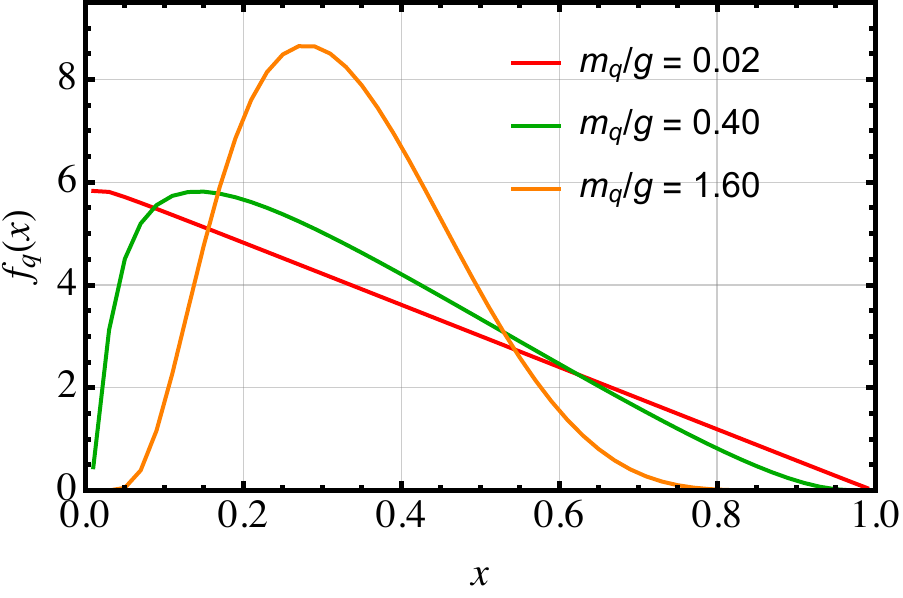}
    }
    \caption{Ground state baryon's PDF for $N$=3 and for different values of $m_q/g$. Solid lines denote the result of our calculation from minimizing the free energy of the baryon; we are not aware of any existing results in the literature to compare against, except in the $m_q/g \rightarrow 0$ limit (red curve), where we reproduce the existing results \cite{Hornbostel:1988fb}, see Eq.~\eqref{eq:pdfbaryonmq0}.  }
    \label{fig:Baryonpdf}
\end{figure}

For the case of $m_q \neq 0$, our results for the mass spectrum are shown in the left panel of Fig.~\ref{fig:Baryonmass}. 
As before, we find good agreement with the existing literature \cite{Hornbostel:1988fb} in the large quark mass limit. 

As $m_q/g$ approaches zero, the agreement between our conjectured free energy result and the Schr\"{o}dinger equation worsens, i.e. that the system is not well simulated as a thermal ensemble for low $m_q$. This underlines the need for further work on refining our free energy functional;
we will discuss some plausible ways for this in the conclusion. 

We also show the PDF derived from our calculation in Fig.~\ref{fig:Baryonpdf}. While the general shape of the PDF looks like what one may have expected, we are not aware of any exact calculation for this quantity to cross check our results.

\section{Summary and Outlook}
\label{sec:conclusion}

This paper is an exploration of how well a hadron in 1+1D is represented as an effective thermal state of free partons, with entanglement entropy of a single parton in a two-parton subsystem playing the role of thermal entropy.

We showed that the mass and the wavefunction (PDF) of hadrons for gauge theories in 1+1D could be \textit{derived} by postulating an emergent minimum free energy principle. This principle, in turn is motivated by the observation that all QFTs have local interactions, which leads to entanglement in momentum space.

In our study of models in 1+1D, we decomposed the Hilbert space into the Fock space of valence quarks; this can be done rigorously in 1+1D since contributions from higher Fock states is suppressed. At the same time, there is no equivalent DGLAP evolution and gluons are not propagating degrees of freedom, further simplifying our calculation. 
We used the entanglement entropy between two partons of a fixed momentum two-parton system and the kinetic energy of the free partons on the light cone in our free energy ansatz. 

The free energy principle can be considered a systematic expansion about the infinite mass limit. In 1+1D theories this is the limit where the system is weakly coupled. 
However, we see that the final PDF ansatz gives astonishingly accurate results even when we enter the strongly coupled region, i.e. when the binding energy forms an $\mathcal{O}(1)$ fraction of the bound state mass. 
In the limit of zero parton mass, our free energy ansatz coincides with negative of entanglement entropy of a single parton. Hence, in this limit the bound states follow a maximum entropy principle.  

This observation holds across distinct models and bound states of models in 1+1D. 
This leads us to conjecture that this general emergent principle, which does not rely on the details of any theory, is successful in capturing the underlying dynamics of bound states of any confining theory. 
We hope this new principle could be systematically generalized to more complicated theories, in particular in higher dimensions.

We can also interpret our proposal as a standard variational method for finding eigenfunctions and eigenvalues of a hadron's Hamiltonian. 
The template functions used in this variational method minimize the free energy of a weekly interacting gas of particles, subject to known constraints on their quantum numbers. 
In our proposed free energy ansatz we have introduced an intrinsic temperature for the hadron. We have calculated its value by postulating that it minimizes the Hamiltonian of the bound state, i.e. it is the optimization parameter in our variational method. 
Further investigation of a physical interpretation of this quantity is left for the future. 

Our work can be extended in many other interesting directions as well. 
We only focused on the ground state hadrons of models in 1+1D. We can repeat this analysis for states with different quantum numbers. Naturally, the sum rules change for that system and the intrinsic temperature could also change. 
Furthermore, our results can be extended to the case of multiple flavors of partons; while this has been studied for the case of mesons \cite{tHooft:1974pnl}, we are not aware of such a study for baryons.

The obvious elephant in the room is the limit of finite but small parton mass. At present we cannot reproduce the existing results obtained by numerically solving light-cone Schr\"{o}dinger equations. 
This clearly means that our proposed free energy is not the correct function to minimize for small masses, which in turn means that a thermal density matrix simply does not work in this regime. 
But given that the maximum entropy principle works at zero mass suggests that some variation of our principle should be plausible when we systematically move away from the zero mass limit.

Ultimately we want to extend our analysis for gauge theories in higher dimensions. 
The natural starting point for venturing into higher dimensions is the heavy quark limit; this is motivated by our principle's success in that limit and known results about hadrons effective Hamiltonian in this regime (from effective theories such as Heavy Quark Effective Theory \cite{Isgur:1989vq,Georgi:1990um,Isgur:1990yhj,Isgur:1991wq}).
One challenge we face here is the issue that the PDF evolves with a renormalization scale. 
While our approach allows us to calculate bound states wavefunction, making a connection with the expectation of the twist two operator corresponding to PDFs will be more challenging in higher dimensions, primarily owing to non-perturbative evolution effects. 

What we would be interested in is whether a general principle can be found which can describe this evolution. 
At the same time, gluons now become propagating degrees of freedom and must be incorporated in any minimization principle. 
Furthermore, in higher dimensions particles will have spins, adding to the number of degrees of freedom and further complicating our numerical calculation. Finally, in higher dimensions the hadron Hilbert space decomposes into valence and unsuppressed sea quarks' Fock space, further increasing the dimension of the Hilbert space.

Nevertheless, the fact that we could derive properties of bound states of various confining models (Schwinger, 't Hooft, and SU($N$) with finite $N$) in 1+1D to within a few percent, suggests our conjectured first principle could be universally applied to accurately derive properties of other confining theories even in higher dimensions. Thus, despite the challenges mentioned above, we believe it is worthwhile trying to generalize our principle to higher dimensions.

If our results are successfully extended to theories in higher dimensions, it is worth trying to make predictions about other properties of hadrons such as fragmentation functions. 
As another phenomenological application of our principle, we can study the form factors of confining dark sector hadrons that could enter various early universe or direct detection observables. We leave all these studies, as well as many other phenomenological directions, for future works.

\section*{Acknowledgment}

We thank Tim Cohen, Matthew Reece, and David Shih for helpful discussions. The work of PA is supported in part by the U.S. Department of Energy under Grant Number DE-SC0011640. 
VV is supported by startup funds from the University of South Dakota.

\afterpage{\clearpage}

\bibliographystyle{utphys}
\bibliography{bib}

\end{document}